\documentclass[11pt,a4paper]{article}

\usepackage[T1]{fontenc}
\usepackage[utf8]{inputenc}
\usepackage[margin=2.5cm]{geometry}
\usepackage{amsmath,amssymb,mathtools}
\usepackage{graphicx}
\usepackage{bm}
\usepackage{microtype}
\usepackage[numbers,compress]{natbib}
\usepackage[colorlinks=true,citecolor=blue,urlcolor=blue,linkcolor=blue]{hyperref}
\usepackage{cleveref}
\usepackage{xcolor}

\AddToHook{cmd/appendix/before}{
    
    \crefalias{section}{appendix}
    \crefalias{subsection}{appendix}
}

\crefformat{equation}{#2Eq.~(#1)#3}
\crefformat{appendix}{#2App.~#1#3}
\crefformat{section}{#2Sec.~#1#3}
\crefformat{table}{#2Tab.~#1#3}
\crefformat{figure}{#2Fig.~#1#3}

\crefmultiformat{section}{#2Secs.~#1#3}{ and~#2#1#3}{, #2#1#3}{ and~#2#1#3}
\crefmultiformat{appendix}{#2Apps.~#1#3}{ and~#2#1#3}{, #2#1#3}{ and~#2#1#3}
\crefmultiformat{equation}{#2Eqs.~(#1)#3}{ and~#2(#1)#3}{, #2(#1)#3}{ and~#2(#1)#3}
\crefmultiformat{figure}{#2Figs.~#1#3}{ and~#2#1#3}{, #2#1#3}{ and~#2#1#3}
\crefmultiformat{table}{#2Tabs.~#1#3}{ and~#2#1#3}{, #2#1#3}{ and~#2#1#3}
\crefrangeformat{equation}{#3Eqs.~(#1)#4--#5(#2)#6}
\crefrangeformat{figure}{#3Figs.~#1#4--#5#2#6}

\crefrangeformat{equation}{#3Eqs.~(#1)#4 to #5(#2)#6}

\Crefformat{figure}{#2Fig.~#1#3}

\newcommand{\vect}[1]{\boldsymbol{#1}}
\newcommand{\MeV}{\;\text{MeV}}

\newcommand{\llambda}{{\lambda\Lambda}}
\begin{document}

\title{A doubly critical point in the color-superconducting regime of the RG-consistent NJL model}

\author{Hosein Gholami\thanks{\texttt{mohammadhossein.gholami@tu-darmstadt.de}}
        \and Marco Hofmann\thanks{\texttt{marco.hofmann.physics@protonmail.com}}
        \and Michael Buballa\thanks{\texttt{michael.buballa@tu-darmstadt.de}}}
 \date{}

\maketitle

\begin{center}
\begin{minipage}{0.9\textwidth}
\small
\textit{Technische Universit\"at Darmstadt, Fachbereich Physik, Institut f\"ur Kernphysik,
Theoriezentrum, Schlossgartenstr.~2, D-64289 Darmstadt, Germany}
\end{minipage}
\end{center}

\begin{abstract}

\noindent
Nambu--Jona-Lasinio (NJL) models of color-superconducting quark matter suffer from
cutoff artifacts once temperature or quark chemical potential become comparable
to the model cutoff. 
These artifacts can be removed employing a renormalization-group (RG) consistent regularization scheme. In this article 
we perform a systematic study of neutral beta-equilibrated quark matter within the three-flavor NJL model with RG consistent regularization. 
Varying the 
coupling constant in the scalar diquark 
channel, we map out the phase diagram in the plane of chemical potential and temperature, 
and identify the gapless domains. We particularly focus on the melting pattern of the color-flavor locked (CFL) phase. 
At 
larger couplings
the CFL phase melts through a so-called $d$SC phase, as expected from leading-order Ginzburg--Landau analyses. Lowering the coupling,
the phase structure becomes markedly richer: While at 
large densities
the CFL phase still melts through a $dSC$ phase, we find a $uSC$ phase at lower chemical potential. These phases, $uSC$ and $dSC$ meet at a \textit{doubly critical point} whose
existence had been anticipated long ago but was never demonstrated explicitly in a model.

\end{abstract}

% =====================================================================
\section{Introduction}
% =====================================================================
Nambu--Jona-Lasinio (NJL)-type models are widely used to study strong-interaction matter at
moderately high density, a region which is phenomenologically interesting for compact-star
physics but not accessible by lattice QCD calculations or perturbation theory. They are
particularly well suited to investigate chiral symmetry breaking and color superconductivity
(CSC) in a single, consistent framework; see Ref.~\cite{Buballa:2003qv} for a review. As shown
in Ref.~\cite{Ruester:2005jc}, the phase
structure of the NJL model in three-flavor, locally neutral quark matter is rich, including
several CSC phases. At intermediate densities, matter is in a two-flavor color-superconducting
(2SC) phase in which up and down quarks of two colors (red and green) are paired, whereas at
high densities the color-flavor-locked (CFL) phase, in which quarks of all flavors and colors
are paired, is preferred. In addition, there can appear other color-superconducting phases such
as the $d$SC phase ($ud$ and $ds$ pairing) and the $u$SC phase ($ud$ and $us$ pairing).

Despite all this, the NJL model is a non-renormalizable theory. In the conventional approach,
integrals are regularized with a single 3d momentum cutoff $\Lambda \simeq 600\,$MeV, which is
fitted to vacuum observables. This leads to cutoff artifacts as soon as the quark chemical
potential $\mu$ or the temperature $T$ become comparable in size to the cutoff. These
well-known artifacts include: (i) the diquark gaps decrease with increasing quark chemical
potential and eventually vanish \cite{Farias:2005cr}; (ii) the speed of sound becomes acausal
\cite{Pasqualotto:2023hho}
and diverges;
(iii) the critical temperature $T_c$ decreases with increasing chemical potential; and
(iv) the melting pattern of the CFL phase disagrees with Ginzburg--Landau (GL) theory
\cite{Iida:2003cc,Iida:2004cj}. In Refs.~\cite{Gholami:2024diy,Gholami:2025guq} we showed that all four are removed by
imposing renormalization-group (RG) consistency \cite{Braun:2018svj} on the three-flavor NJL
model with color superconductivity. The corresponding treatment of the two-flavor
Quark--Meson--Diquark model is given in Ref.~\cite{Gholami:2025afm}, and astrophysical
applications in Refs.~\cite{Gholami:2024ety,Christian:2025dhe,Kunkel:2026wvz}.

The artifact (iv) is the most subtle of the four and the main focus of the present work. 
When CFL matter is heated up, the three diquark gaps related to $d$-$s$, $u$-$s$ and $u$-$d$ pairing, respectively, melt successively, leading to two intermediate phases before finally reaching the completely unpaired ``normal conducting'' (NQ) phase. 
The $u$-$d$ pairs are the most stable ones, so the last color-superconducting phase below the NQ phase is a 2SC phase. 
However, whether the $u$-$s$ pairs or the $d$-$s$ pairs break up first, i.e., whether in the region between CFL and 2SC phase there is a $d$SC or a $u$SC phase,
has been subject of active research
\cite{Fukushima:2004zq,Iida:2003cc,Iida:2004cj,Abuki:2005ms}.
A Ginzburg-Landau (GL) analysis at small $M_s^2/\mu^2$, where $M_s$ is the dressed strange-quark mass and $\mu$ is the chemical potential, predicts an intermediate $d$SC phase
\cite{Iida:2003cc,Iida:2004cj},
whereas in an NJL-model study with conventional regularization a $u$SC window was found \cite{Ruester:2005jc,Abuki:2005ms}. 
This puzzle was basically resolved by the observation that the RG-consistent regularization of the same model yields a $d$SC phase  \cite{Gholami:2024diy}.
On the other hand, it was shown in Ref.~\cite{Abuki:2005ms} that, due to higher-order corrections in the 
$M_s^2/\mu^2$, there actually could be a $u$SC phase at lower chemical potential. 
Indeed, in the RG consistent analysis of Ref.~\cite{Gholami:2024diy}, precursors of a second-order phase transition to a $u$SC phase were seen in the CFL phase at low $\mu$, 
but this transition was eventually prevented by an intervening first-order CFL~$\to$~2SC transition.  
Therefore the question arises whether a $u$SC phase can be made to appear by varying the diquark coupling strength. 
In this case a \textit{doubly critical point} (DCP) \cite{Fukushima:2004zq,Abuki:2005ms,Fukushima:2005fh} must exist at which the $u$SC phase connects to the $d$SC phase, the latter always being present at high densities independently of the parameters.

In this contribution we close that gap. To this end we perform a scan of the diquark coupling $G_D$ down to values at which the onset of color superconductivity
is pushed to larger $\mu$. 
We also resolve the domains of \textit{gapless} CSC, i.e., regions in the phase diagram where quarks are paired but there is no gap in the quasiparticle spectrum \cite{Shovkovy:2003uu}.
We eventually find the anticipated DCP in a region where both the $u$SC phase and the $d$SC phase as well as the neighboring CFL phase are all gapless.

% =====================================================================
\section{Model and RG-consistent minimal scheme}
\label{sec:model}
% =====================================================================

We work with the three-flavor NJL model of Ref.~\cite{Ruester:2005jc}, which is defined by
the Lagrangian
\begin{equation}\label{eq:Lagrangian}
\mathcal{L}=\bar{\psi}\left(i\gamma^\mu\partial_\mu+\gamma^0\hat{\mu}-\hat{m}\right)\psi
+\mathcal{L}_{\bar{q}q}+\mathcal{L}_{qq}.
\end{equation}
Here $\psi$ denotes a quark spinor field with three flavor ($u$, $d$, $s$) and three color
($r$, $g$, $b$) degrees of freedom, $\hat{m}=\text{diag}_f(m_u,m_d,m_s)$ is the matrix of
the bare quark masses, and $\hat{\mu}$ is a matrix of chemical potentials in flavor and
color space, which will be specified below. The quarks interact via the quark--antiquark
vertices
\begin{equation}\label{eq:Lqbarq}
\mathcal{L}_{\bar{q}q}=
G_S\sum_{a=0}^{8}\left[(\bar{\psi}\tau_a\psi)^2+(\bar{\psi}i\gamma_5\tau_a\psi)^2\right]
-K\left[\text{det}_f\big(\bar{\psi}(1+\gamma_5)\psi\big)
+\text{det}_f\big(\bar{\psi}(1-\gamma_5)\psi\big)\right],
\end{equation}
consisting of a $U(3)_L\times U(3)_R$ symmetric four-point interaction with coupling
constant $G_S$, where $\tau_a$ denote the Gell-Mann matrices in flavor space, complemented
by $\tau_0=\sqrt{2/3}\,\mathbf{1}_f$, and the six-point Kobayashi-Maskawa-'t Hooft interaction with coupling
constant $K$, which is $SU(3)_L\times SU(3)_R$ symmetric but explicitly breaks the axial
$U(1)_A$ symmetry. In addition, the quark--quark interaction
\begin{equation}\label{eq:Lqq}
\mathcal{L}_{qq}=
G_D\sum_{\gamma,c}
\left(\bar{\psi}^a_\alpha\, i\gamma_5\,\epsilon^{\alpha\beta\gamma}\epsilon_{abc}\,
(\psi_C)^b_\beta\right)
\left((\bar{\psi}_C)^r_\rho\, i\gamma_5\,\epsilon^{\rho\sigma\gamma}\epsilon_{rsc}\,
\psi^s_\sigma\right),
\end{equation}
with flavor indices $\alpha,\beta,\ldots$, color indices $a,b,\ldots$, and the
charge-conjugate spinors $\psi_C=C\bar{\psi}^T$, where $C=i\gamma^2\gamma^0$, describes
scalar diquark pairing in the color and flavor antitriplet channels, which are the relevant
channels for both 2SC and CFL pairing.

We treat this model in the mean-field (Hartree) approximation, allowing for nonvanishing
expectation values of the three quark--antiquark condensates
\begin{equation}\label{eq:phidef}
\phi_f=\langle\bar{\psi}_f\psi_f\rangle, \qquad f=u,d,s,
\end{equation}
which are related to the dressed quark masses,
$M_\alpha=m_\alpha-4G_S\phi_\alpha+2K\phi_\beta\phi_\gamma$, with $(\alpha,\beta,\gamma)$
being any permutation of $(u,d,s)$, and of the three diquark condensates, corresponding to
the gaps
\begin{equation}\label{eq:DeltaA}
\Delta_A=-2G_D\,\langle\bar{\psi}^a_\alpha\, i\gamma_5\,
\epsilon^{\alpha\beta A}\epsilon_{abA}\,(\psi_C)^b_\beta\rangle,
\qquad A=1,2,3,
\end{equation}
which are the order parameters of color superconductivity. The gap $\Delta_A$ describes the
pairing of quarks of the two flavors and two colors not equal to $A$; we therefore also
label the gaps by the flavors they pair, i.e., $\Delta_{ud}\equiv\Delta_3$,
$\Delta_{us}\equiv\Delta_2$ and $\Delta_{ds}\equiv\Delta_1$. For convenience, we collect
the six condensates in a single field
$\vect\chi=(\phi_u,\phi_d,\phi_s,\Delta_{ud},\Delta_{us},\Delta_{ds})$.
In this approximation the effective potential per volume takes the general form
\begin{equation}\label{eq:Omega_eff}
\Omega_{\text{eff}}(T,\vect{\mu};\vect\chi)
=\mathbb{L}(T,\vect{\mu};\vect\chi)+\mathcal{V}(\vect\chi),
\end{equation}
with the condensate potential
\begin{equation}\label{eq:V_def}
\mathcal{V}(\vect\chi)=2G_{S}(\phi_u^2+\phi_d^2+\phi_s^2)-4K\phi_u\phi_d\phi_s
+\frac{1}{4G_D}\sum_{A=1}^3\vert\Delta_A\vert^2 ,
\end{equation}
and the fermionic loop part
\begin{equation}\label{eq:Ldef}
\mathbb{L}(T,\vect{\mu};\vect\chi)=-\int\!\frac{d^3p}{(2\pi)^3}\,
\mathcal{A}(T,\vect{\mu};\vect\chi),
\end{equation}
where $\mathcal{A}$, which also depends on the modulus $p$ of the three-momentum, is
obtained from the trace of the logarithm of the inverse dressed quark propagator in flavor,
color, Dirac and Nambu--Gorkov space. Details of the block decomposition of the propagator
and of the resulting dispersion relations are given in
Refs.~\cite{Ruester:2005jc,Gholami:2024diy}. For given temperature and chemical potentials,
the physical values of the condensates are those which minimize the effective potential,
i.e., the solutions of the gap equations
$\partial\Omega_{\text{eff}}/\partial\vect\chi=0$.

Following Ref.~\cite{Ruester:2005jc} we consider the set of chemical potentials appropriate
for cold compact stars. Because of weak decays, quark flavor is not conserved, and the chemical
potential matrix reads
\begin{equation}\label{eq:mu_matrix}
\hat{\mu}^{\alpha\beta}_{ab}=(\mu\delta^{\alpha\beta}+\mu_Q Q^{\alpha\beta})\delta_{ab}
+[\mu_3 (\lambda_3)_{ab}+ \mu_8 (\lambda_8)_{ab}]\delta^{\alpha\beta},
\end{equation}
with the electric charge operator $Q=\text{diag}_f(2/3,-1/3,-1/3)$ and the third and eighth
Gell-Mann matrices in color space, $\lambda_3$ and $\lambda_8$, respectively. Adding a free Fermi gas of electrons and muons with
$\mu_e=\mu_\mu=-\mu_Q$, we impose electric and color neutrality,
\begin{equation}\label{eq:neutrality}
\bigg[\frac{\partial \Omega_\text{eff}^\text{total}}{\partial \mu_Q}\bigg]_{\mu_Q=\bar{\mu}_Q}
=\bigg[\frac{\partial \Omega_\text{eff}^\text{total}}{\partial \mu_3}\bigg]_{\mu_3=\bar{\mu}_3}
=\bigg[\frac{\partial \Omega_\text{eff}^\text{total}}{\partial \mu_8}\bigg]_{\mu_8=\bar{\mu}_8}=0,
\end{equation}
leaving the quark chemical potential $\mu$ as the single independent one. 
Here $\Omega_\text{eff}^\text{total}$ is the sum of $\Omega_\text{eff}$ from \cref{eq:Omega_eff} and the lepton free energy density.
Where several
solutions of the gap equations and \cref{eq:neutrality} coexist, the physical one is that with
the largest pressure $P=-\Omega^{\text{total}}(T,\mu)$.

The momentum integral in the loop term \cref{eq:Ldef} is ultraviolet divergent and therefore
needs to be regularized. In vacuum this is done by restricting the integral to momenta below
a sharp three-momentum cutoff $\Lambda$, which, together with the other model parameters, is
fitted to vacuum observables. Keeping this cutoff for the medium parts as well, as done in
the conventional approach, leads to the cutoff artifacts (i)--(iv) mentioned in the
introduction as soon as $\mu$ or $T$ become comparable to $\Lambda$. These artifacts can be
removed by demanding RG consistency, i.e., by requiring that the full quantum effective
action must not depend on the UV scale at which the classical action is initialized
\cite{Braun:2018svj}. In practice this means that the cutoff $\Lambda$ is kept only for the
divergent vacuum part of $\mathbb{L}$, while the medium part is integrated up to a much
higher scale $\llambda \gg \mu,T,\Lambda$, where $\lambda$ measures the UV scale in units
of the model cutoff and could ideally be sent to infinity. In normal-conducting quark
matter this is straightforward, since there the medium contributions are UV finite. In the
presence of CSC condensates, however, the medium contributions are divergent themselves,
behaving as
$\sim\sum_{\alpha a,\beta b}\mu^2_{\alpha a,\beta b}\,|\Delta_{\alpha a,\beta b}|^2\ln\llambda$, where
$\Delta_{\alpha a,\beta b}$ denotes the gap associated with the pairing of the quarks with
flavor and color $(\alpha a)$ and $(\beta b)$, i.e.,
$\Delta_{ur,dg}=\Delta_{ug,dr}=\Delta_{ud}$ etc., and $\mu_{\alpha a,\beta b}$ is the
average chemical potential of that pair. We can therefore not simply regularize the vacuum
part only and leave the medium part unregularized. Instead, we have to perform additional
subtractions. Specifically, we take the following expression for the effective potential:
\begin{equation}\label{eq:Omega_min}
\Omega_{\text{eff}}(T,\vect{\mu};\vect\chi)
=\mathcal{V}(\vect\chi)
-\frac{1}{2\pi^2}\left(\int_0^{\llambda}\!dp\,p^2\,\mathcal{A}(T,\vect{\mu};\vect\chi)
-\int_{\Lambda}^{\llambda}\!dp\,p^2\,\mathcal{A}_{\text{vac}}(\vect\chi)\right)
-\delta\mathcal{L}^{\text{(min)}} ,
\end{equation}
where $\mathcal{A}_{\text{vac}}(\vect\chi)\equiv\mathcal{A}(0,\vect{0};\vect\chi)$ is the
vacuum part of $\mathcal{A}$. The subtraction of $\mathcal{A}_{\text{vac}}$ on the interval
$[\Lambda,\llambda]$ ensures that the vacuum contribution is regularized by the model
cutoff $\Lambda$, while the medium part is integrated up to the scale $\llambda$; the
counterterm $\delta\mathcal{L}^{\text{(min)}}$ subtracts the divergent medium
contributions.

This subtraction is not unique: only the divergent part of the counterterm is fixed, and
finite terms may be subtracted along with it, corresponding to different RG-consistent
regularization schemes, which are discussed in
Refs.~\cite{Gholami:2024diy,Gholami:2026qpi}. Throughout this work we use the
\emph{minimal} scheme, in which only the terms required for the cancellation of the
divergence are subtracted. It is defined by the counterterm
\begin{equation}\label{eq:minimaldef}
\delta\mathcal{L}^{\text{(min)}}=-\frac{1}{2\pi^2}\int_{\Lambda}^{\llambda} dp\,p^2
\sum\frac{1}{4}\,\mu_{\alpha a,\beta b}^2\Delta_{\alpha a,\beta b}^2
\left[\frac{\partial^4}{\partial\Delta_{\alpha a,\beta b}^2\partial\mu_{\alpha a,\beta b}^2}
\mathcal{A}(0,\vect{\mu},\vect\chi)\right]_{\vect\mu=\vect M=\vect\Delta=0},
\end{equation}
which retains only the $\Delta_{\alpha a,\beta b}^2$ terms actually needed to cancel the
divergence, and can be evaluated analytically,
\begin{equation}\label{eq:minimalexpl}
\delta\mathcal{L}^{\text{(min)}}=-\frac{1}{\pi^2}\ln\lambda\,\bigg(
(\mu_{ur,dg}^2+\mu_{ug,dr}^2)\Delta_{ud}^2
+(\mu_{ur,sb}^2+\mu_{ub,sr}^2)\Delta_{us}^2
+(\mu_{dg,sb}^2+\mu_{sg,db}^2)\Delta_{ds}^2\bigg).
\end{equation}
We restrict the present study to the minimal scheme, for two related reasons. The first is that
it admits a field-theoretical interpretation that the other schemes lack: the counterterm
\cref{eq:minimalexpl} is precisely what one obtains from a \emph{wave-function renormalization}
of the diquark fields, that is, from writing $Z_{\Delta_A}=1+\delta Z_{\Delta_A}$ and demanding
that the scale-dependent renormalization constant vanishes at the model cutoff. In the NJL model
the diquark kinetic term is generated only dynamically, so this remains a motivation rather than
a derivation; but in the Quark--Meson--Diquark model the kinetic term is present from the outset,
and there the same construction can be carried through explicitly
\cite{Gholami:2024diy,Gholami:2025afm}.

The second reason is that the minimal scheme reproduces the weak-coupling benchmarks. 
In contrast to the other schemes introduced in Ref.~\cite{Gholami:2024diy}, it
preserves the BCS relation between the critical temperature and the zero-temperature gap for
symmetric two-flavor matter, and the corresponding CFL relation for three flavors \cite{Gholami:2025afm}; the explicit ratios are given in
\cref{eq:bcsratios} below. Since the present work relies on the behavior of the gap at large
chemical potential, a scheme that respects these ratios is the more faithful qualitative guide
to the phase diagram, and we adopt it throughout.

We employ the full parameter set of Ref.~\cite{Ruester:2005jc}: $\Lambda=602.3\MeV$,
$G_S\Lambda^2=1.835$, $K\Lambda^5=12.36$, and bare quark masses $m_{u,d}=5.5\MeV$,
$m_s=140.7\MeV$, fitted to the meson spectrum in vacuum \cite{Rehberg:1995kh}. The UV scale is
set to $\lambda=10$, which is large enough to eliminate cutoff artifacts
\cite{Gholami:2024diy}. The diquark coupling $G_D$ is the parameter we vary.

\subsection{Asymptotic scales and what controls them}
\label{sec:asymptotics}

The interpretation of the results below rests on the behavior of the model at asymptotically
large chemical potential, which in the minimal scheme is known in closed form. In the symmetric,
massless limit the zero-temperature diquark gaps saturate at
\begin{equation}\label{eq:asympgaps}
\bar\Delta^{\text{(min.)}}_{\text{asymp},\,ud} = 2\,\Lambda\,e^{-3/2},
\qquad
\bar\Delta^{\text{(min.)}}_{\text{asymp},\,uds} = 2^{2/3}\,\Lambda\,e^{-3/2}
= 2^{-1/3}\,\bar\Delta^{\text{(min.)}}_{\text{asymp},\,ud},
\end{equation}
for 2SC and symmetric CFL pairing respectively, while the critical temperature saturates at
\begin{equation}\label{eq:asympTc}
T_c(\mu \to \infty) = \frac{2}{\pi}\,\Lambda\,e^{\gamma_E-3/2},
\end{equation}
with $\gamma_E \approx 0.577$ the Euler--Mascheroni constant. Derivations are given in
Refs.~\cite{Gholami:2025afm,Gholami:2025guq,Gholami:2026qpi}. Although
\cref{eq:asympgaps,eq:asympTc} are derived in the symmetric massless limit, they control the
full neutral calculation as well, because that limit is precisely what the neutral model
approaches as $\mu\to\infty$: the strange quark mass goes to zero, the flavor densities equalize, and
the neutralizing chemical potentials become irrelevant compared with $\mu$
\cite{Gholami:2024diy}. For our parameters,
\begin{equation}\label{eq:asympnumbers}
\bar\Delta^{\text{(min.)}}_{\text{asymp},\,ud} \approx 268.8\MeV, \qquad
\bar\Delta^{\text{(min.)}}_{\text{asymp},\,uds} \approx 213.3\MeV, \qquad
T_c(\mu\to\infty) \approx 152.4\MeV .
\end{equation}

Three features of \cref{eq:asympgaps,eq:asympTc} organize everything that follows.

First, \emph{the diquark coupling does not appear}. $G_D$ controls where color superconductivity
sets in and how large the gaps are at moderate density, but it drops out of the $\mu\to\infty$
limit entirely: the asymptotic scales are fixed by the vacuum cutoff scale $\Lambda$ alone.

Second, this fixes the asymptotic value of the diquark gap while leaving the chemical potential where it becomes nonzero free.
Lowering $G_D$ pushes the onset of pairing to larger $\mu$ without changing the gap's asymptotic value. At a fixed value of $\mu$, a lower $G_D$ leads to a smaller gap.
Therefore, since smaller gaps are more sensitive to pairing stress caused by the Fermi surface mismatch, which exists due to requiring electric neutrality and which is rather independent of $G_D$, we expect the emergence  of more interesting phase structures, including large regions of gapless phases, when $G_D$ is lowered. 

Third, the ratios are more robust than the scales themselves. The cutoff cancels in
\begin{equation}\label{eq:bcsratios}
\lim_{\mu\to\infty}\frac{T_c}{\bar\Delta_{ud}} = \frac{e^{\gamma_E}}{\pi} \approx 0.567,
\qquad
\lim_{\mu\to\infty}\frac{T_c}{\bar\Delta_{uds}} = 2^{1/3}\,\frac{e^{\gamma_E}}{\pi} \approx 0.714,
\end{equation}
which are independent of $G_D$ and $\Lambda$, and are equal to the BCS value and its CFL counterpart from weak-coupling QCD \cite{Alford:2007xm,Schmitt:2002sc}. They are thus parameter-free predictions of the RG-consistent minimal scheme, whereas the absolute values in \cref{eq:asympnumbers} inherit the vacuum scale $\Lambda$.

We add one caveat. That the gaps saturate at all is a feature of the model rather than of QCD,
where the weak-coupling gap continues to grow with $\mu$ as the running coupling decreases \cite{Son:1998uk}.
The RG-consistent treatment removes the spurious \emph{decrease} of the gap produced by the
cutoff, but it does not manufacture the QCD rise; what it delivers is a finite, controlled high-density limit in which the model can be compared with weak-coupling expectations at
all \cite{Gholami:2024diy}.

% =====================================================================
\section{Gapless domains in the neutral phase diagram}
\label{sec:gapless}
% =====================================================================

In Ref.~\cite{Gholami:2024diy} we presented the RG consistent phase diagram of neutral quark matter at the benchmark coupling
$G_D=G_{S}$. Comparing it with the result of Ref.~\cite{Ruester:2005jc}, which was obtained in the same model but employing standard cutoff regularization, we identified several regularization artifacts in the latter which were removed by the RG consistent treatment. However, unlike in Ref.~\cite{Ruester:2005jc}, we did not distinguish between gapped and gapless color-superconducting phases in Ref.~\cite{Gholami:2024diy}. 
This distinction is important, since gapless phases are potentially unstable \cite{Huang:2004bg}, at least at low temperature.
As we have argued above, we expect gapless phases to become even more relevant at lower values of $G_D$. 
Moreover, as we will show, the DCP must be located in a gapless area. 
Therefore we begin our analysis to resolve the gapless regimes.

Gapless color superconductivity \cite{Shovkovy:2003uu,Alford:2003fq,Ruester:2004eg} is a direct
consequence
of the neutrality constraints \cref{eq:neutrality}. In the 2SC phase, $\beta$ equilibrium and
electric neutrality require a nonzero electric chemical potential, which splits the $u$ and $d$
Fermi surfaces by a mismatch $\delta\mu$. Pairing is then a competition between this mismatch
and the gap: as long as $\delta\mu < \Delta_{ud}$, all paired quasiparticles remain gapped,
whereas for $\delta\mu > \Delta_{ud}$ the lower quasiparticle branch develops zeros at two
momenta and the excitation spectrum becomes \emph{gapless} without the condensate vanishing.
The resulting phase -- g2SC in this case -- carries the same symmetry breaking pattern as its
gapped counterpart. At nonzero temperature the two are therefore not separated by a genuine
phase transition but by a smooth crossover \cite{Ruester:2005jc}.
The same mechanism operates for the strange-quark pairs,
where the mismatch is controlled in addition by $M_s$, giving rise to g$d$SC, g$u$SC and gCFL
\cite{Alford:2003fq}. Since the mismatches grow as the gaps melt, the gapless regions
generically appear as layers just below the second-order phase boundaries. Across first-order boundaries, on the other hand, the gaps jump
discontinuously, so for instance a gapped CFL phase can be adjacent to a gapped or gapless phase of another color-superconducting pairing, without an intervening gCFL layer.

\begin{figure}[t]
\centering
\begin{minipage}[t]{0.48\textwidth}
\includegraphics[width=\textwidth]{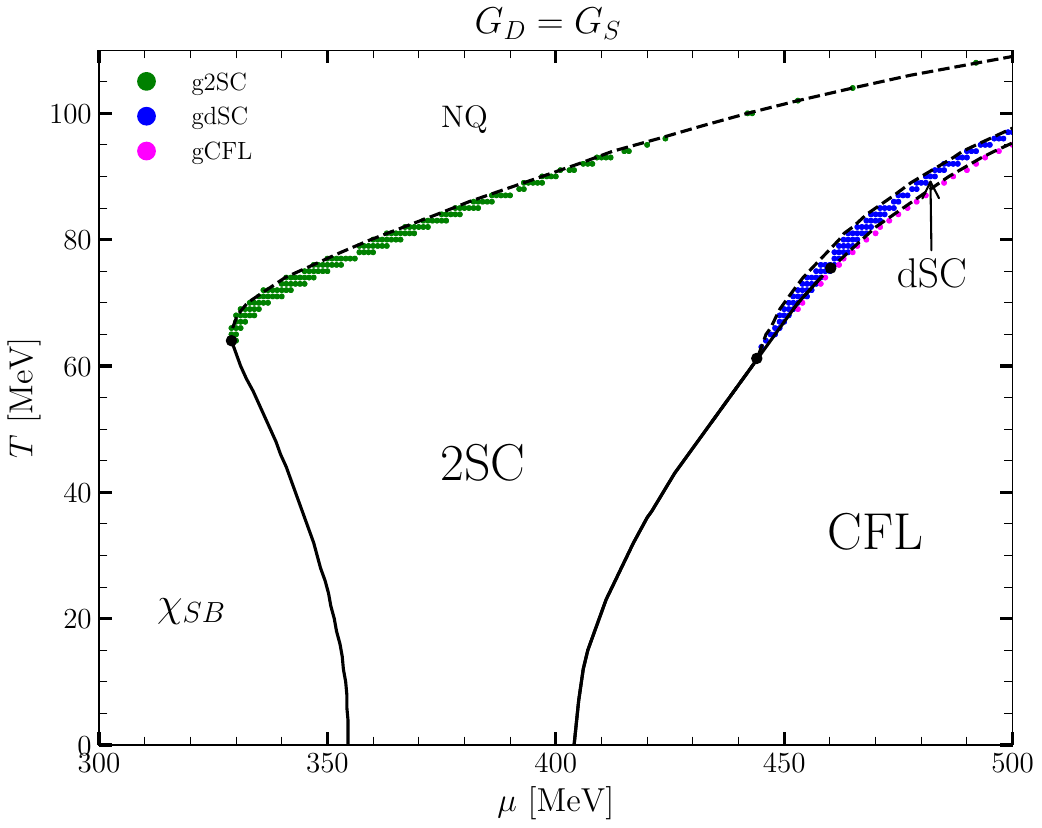}
\end{minipage}\hfill
\begin{minipage}[t]{0.48\textwidth}
\includegraphics[width=\textwidth]{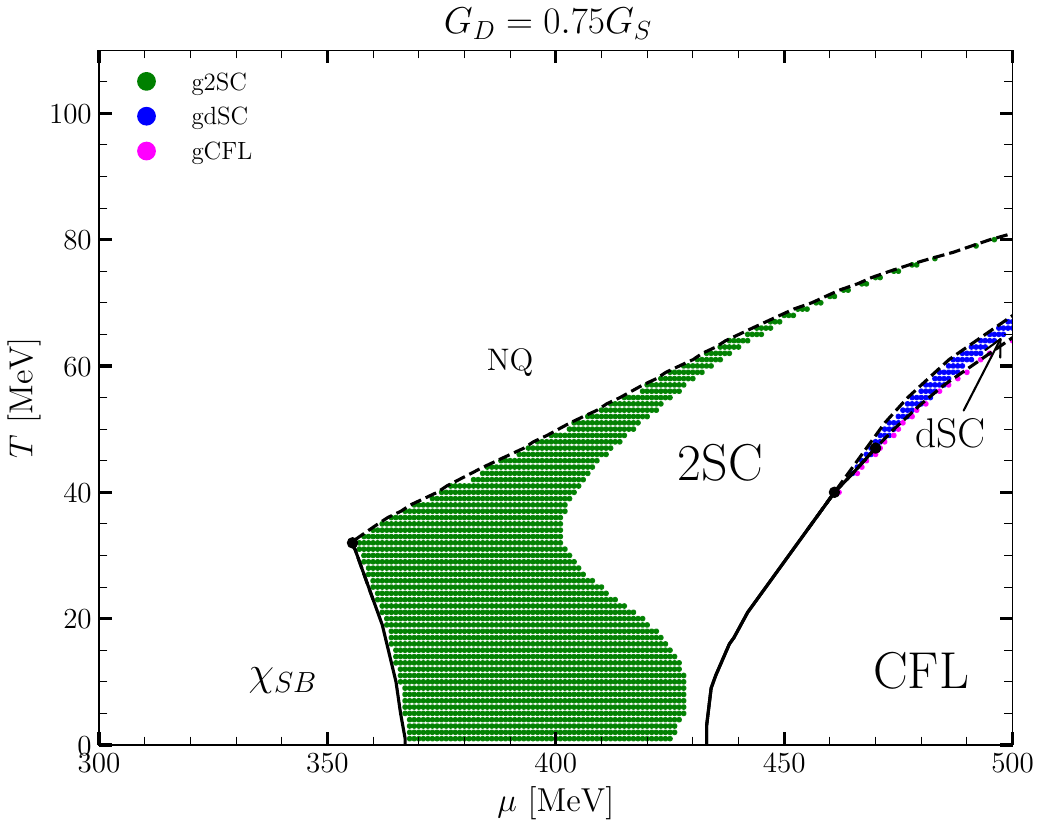}
\end{minipage}
\caption{Phase diagram for neutral quark matter in the RG-consistently regularized NJL model 
(minimal scheme) for two couplings:
$G_D=G_{S}$ (left) and $G_D=\tfrac{3}{4}G_{S}$ (right). Gapless regimes are indicated as hatched areas.
}
\label{fig:pdNJLgaps}
\end{figure}

\Cref{fig:pdNJLgaps} shows the phase diagrams for two different coupling constants $G_D=G_{S}$ and $G_D=\tfrac{3}{4}G_{S}$.
The results can directly be compared with Ref.~\cite{Ruester:2005jc}, where the same parameters have been taken for the model with standard cutoff regularization. 
Following that reference, we also distinguish between two sub-areas of the non-CSC regime: the region where chiral symmetry is spontaneously broken ($\chi$SB) and the region where it is approximately restored (NQ). Gapless regimes are indicated as hatched areas.  
For $G_D=G_{S}$ (left panel) we find relatively narrow, yet infinitely long wedges of gCFL, g2SC, and
g$d$SC, which persist to higher $\mu$ values and lie just below their complete melting along
the second-order phase-boundary lines. Lowering the coupling to $G_D=\tfrac{3}{4}G_{S}$ (right panel)
substantially enlarges the g2SC region. Recall from \cref{sec:asymptotics} that $T_c$ and the
asymptotic zero-temperature gaps are independent of $G_D$. Consequently, the high-$\mu$ part of
the diagram approaches the same asymptotic structure for both couplings beyond the shown region.

The pattern of transitions at fixed $\mu$ can differ between these couplings. At
$G_D=\tfrac{3}{4}G_{S}$ a gapless 2SC phase can already occur at $T=0$, allowing both a direct
g2SC$\to$NQ transition and a two-step 2SC$\to$g2SC$\to$NQ sequence (the latter sequences also
appear for $G_D=G_{S}$). At higher $\mu$, the melting sequences involving gapless phases vary.
If we  restrict ourselves to second-order transitions,
we observe either
\begin{equation*}
\text{CFL}\;\to\;\text{gCFL}\;\to\;\text{$d$SC}\;\to\;\text{g$d$SC}\;\to\;\text{2SC}\;\to\;\text{g2SC}\;\to\;\text{NQ},
\end{equation*}
or
\begin{equation*}
\text{CFL}\;\to\;\text{gCFL}\;\to\;\text{g$d$SC}\;\to\;\text{2SC}\;\to\;\text{g2SC}\;\to\;\text{NQ},
\end{equation*}
for both values of $G_D$.
Thus, in both cases the melting of the CFL phase proceeds via a gapped or gapless $d$SC
phase: for both couplings, the $u$SC phase found in the conventional regularization
\cite{Ruester:2005jc} is replaced by $d$SC, and no $u$SC region is observed anywhere in the
phase diagram. We come back to this important observation in \cref{sec:melting}.

Note that for $G_D=\tfrac{3}{4}G_{S}$, we still find intervals of 2SC and g2SC pairing at $T=0$. This is in contrast to the conventional-cutoff analysis of Ref.~\cite{Ruester:2005jc}, where an NQ layer was found at low temperatures, and (g)2SC phases only exist above $T\gtrsim 10$~MeV. 
This can be understood from the fact that the cutoff effectively weakens the pairing strength. This leads to smaller gaps at $\delta\mu = 0$, making the pairs more fragile when the Fermi surfaces are driven apart by imposing neutrality conditions.
At low $T$, (g)2SC pairing was then prevented completely in the example at hand, and became only possible at higher $T$ with the help of thermal smearing of the Fermi surfaces.
With RG consistent regularization, on the other hand, the 2SC gaps are larger and therefore can partially sustain the pairing stress, even at $T=0$.

The trend with $G_D$ seen in \cref{fig:pdNJLgaps} is best understood at lower temperatures from the criterion
$\delta\mu \gtrless \Delta_{ud}$ for a massless 2SC matter: lowering $G_D$ reduces the diquark gaps at moderate
chemical potentials, leaving more room for gapless pairing and thus explaining the
substantially larger g2SC region at $G_D=\tfrac{3}{4}G_{S}$. At large $\mu$, on the other
hand, gapless pairing is suppressed by two effects. First, the zero-temperature gap grows
towards its asymptotic value \cref{eq:asympgaps}, which is independent of $G_D$
(see \cref{sec:asymptotics}), and thereby overtakes the mismatch, so that the fully gapped
2SC phase is restored; a similar restoration was found in the two-flavor NJL-model study of
Ref.~\cite{Duarte:2018kfd}, employing the regularization scheme of Ref.~\cite{Farias:2005cr},
in which the medium contributions are left unregularized as well. Second, the strange quark
mass falls with increasing $\mu$, so that the neutrality conditions \cref{eq:neutrality}
require a smaller electric chemical potential and the mismatch itself is
reduced.\footnote{In a pure two-flavor model this second effect is absent: there $\delta\mu$
keeps growing with $\mu$ while the gap saturates, so that the g2SC phase eventually
reappears at sufficiently high chemical potential.}

Note that this suppression does not remove the gapless layers along the second-order phase
boundaries: since the gaps vanish continuously there while the Fermi-surface mismatches
remain finite, thin gapless shells persist just below the transition lines at \emph{any}
chemical potential.
What the suppression controls is the extent of the gapless regions at low temperature, deep
inside the paired phases, where it can separate distinct gapless domains from one another.
This is seen most cleanly at $G_D=0.6\,G_{S}$, where a wide band of fully gapped 2SC
separates the g2SC and gCFL domains, as we will see in \cref{sec:dcp}.

% =====================================================================
\section{The CFL melting pattern and the case for a doubly critical point}
\label{sec:melting}
% =====================================================================
The question whether the phase diagram exhibits a $d$SC or a $u$SC phase in the temperature
region above the CFL phase has been a subject of active research in the past decades
\cite{Fukushima:2004zq,Iida:2003cc,Iida:2004cj,Abuki:2005ms}. The authors of
Refs.~\cite{Iida:2003cc,Iida:2004cj} performed a GL expansion of the thermodynamic potential
around the critical temperature $T_{c0}$ of the second-order CFL~$\to$~NQ phase transition
existing in the limit of three massless quark flavors. Taking into account effects of a nonzero
strange mass and charge neutrality to quadratic order in $M_s$, they found that at very high
densities where $M_s \ll \mu$, CFL melts via an intermediate $dSC$ phase.

In Ref.~\cite{Abuki:2005ms}, the expansion was extended to quartic order in
$M_s$.\footnote{The leading- and next-to-leading order $M_s$ corrections are proportional to
$M_s^2/\mu^2 \ln(\mu/T_{c0})$ and $M_s^4/(\mu^2 T_{c0}^2)$, respectively.} The authors found
that, with this additional term, the $d$SC phase could turn into a $u$SC phase at lower
chemical potential. More precisely, they showed that at this order the melting pattern contains an intermediate $d$SC ($u$SC) phase if the squared strange quark mass is less (greater) than
\begin{equation}\label{eq:Mscrit}
M_{s,\text{crit}}^2 = \frac{32\pi^2}{21\zeta(3)}\, T_{c0}^2 \ln(\mu/T_{c0}) .
\end{equation}
Since at asymptotic chemical potential the gaps and, hence, $T_{c0}$ become infinitely large in
QCD, this means that at sufficiently large $\mu$ there must be a $d$SC phase, as already
predicted in Refs.~\cite{Iida:2003cc,Iida:2004cj}. The same conclusion holds in our model even
though its gaps saturate rather than grow, \cref{eq:asympgaps}: with $T_{c0}$ approaching the
constant \cref{eq:asympTc}, the right-hand side of \cref{eq:Mscrit} still diverges
logarithmically with $\mu$, while $M_s \to 0$. It is therefore the vanishing of the strange
quark mass, together with pairing gaps that stay nonzero, that guarantees a $d$SC phase at
sufficiently high density here. At lower chemical potential, on the other
hand, a $u$SC phase might be realized, meaning that the presence of this phase in the
conventional NJL studies is not necessarily unphysical. Qualitatively, the result is also
consistent with the NJL-model-type calculation of Ref.~\cite{Fukushima:2004zq}, where
$M_s^2/\mu$ was treated as a free parameter.

\begin{figure}[t]
\centering
\begin{minipage}[t]{0.32\textwidth}
\includegraphics[width=\textwidth]{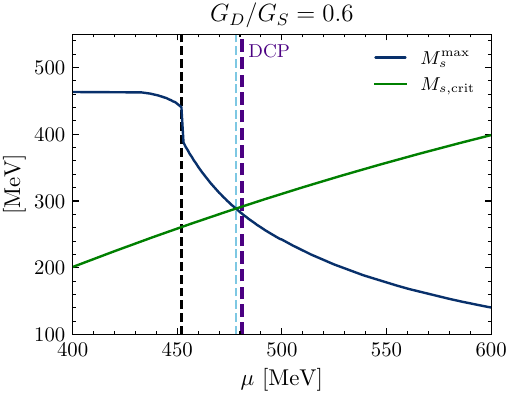}
\end{minipage}\hfill
\begin{minipage}[t]{0.32\textwidth}
\includegraphics[width=\textwidth]{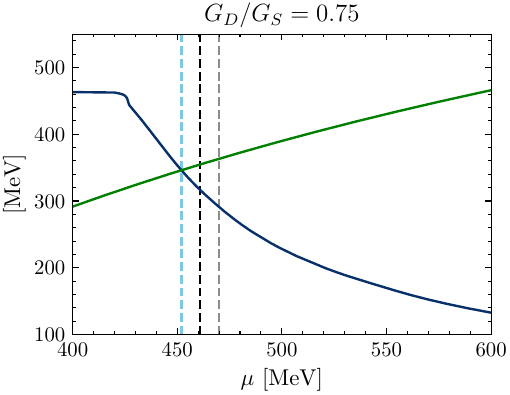}
\end{minipage}\hfill
\begin{minipage}[t]{0.32\textwidth}
\includegraphics[width=\textwidth]{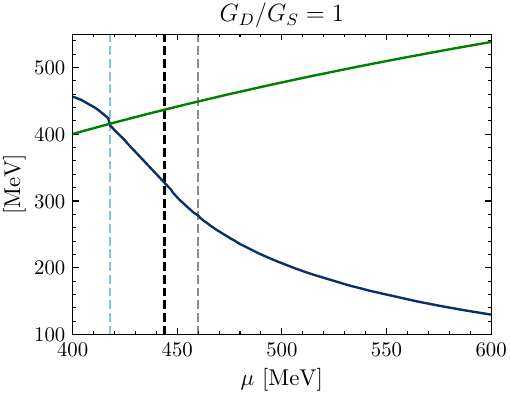}
\end{minipage}
\caption{Maximum strange-quark mass $M_s^{\max}$ of the thermodynamically preferred
solution (dark blue) and the GL threshold $M_{s,\text{crit}}$ from \cref{eq:Mscrit}
(green) as functions of the chemical potential, in the minimal scheme for
$G_D/G_{S}=0.6$ (left), $0.75$ (center) and $1$ (right). The light-blue dashed lines mark
the crossing $M_s^{\max}=M_{s,\text{crit}}$: to its left (right) the GL analysis predicts
a $u$SC ($d$SC) melting pattern. Black and gray dashed lines indicate the critical
endpoints of the first-order CFL phase boundary, taken from the phase diagrams in
\cref{fig:pdNJLgaps,fig:pdNJLmin06gapless}; the GL comparison applies only in the region
of sequential second-order melting beyond them. The purple dashed line in the left panel
marks the DCP at $\mu\approx481\MeV$ (see \cref{sec:dcp}).}
\label{fig:MsCrit}
\end{figure}

A comparison of both sides of 
\cref{eq:Mscrit}
was first carried out quantitatively in Ref.~\cite{Gholami:2024diy}. There
it was also found that the GL criterion predicts a $d$SC pattern even for the conventional
regularization, which nevertheless produces a $u$SC window -- one of the indications that
the latter is a cutoff artifact. In \cref{fig:MsCrit} we perform this comparison for the
minimal scheme employed here, for the three couplings considered in this work. At each
chemical potential, the strange-quark mass of the model is taken as its maximum with
respect to temperature over the thermodynamically preferred solutions, denoted
$M_s^{\max}$, so as to obtain an upper limit, and is confronted with $M_{s,\text{crit}}$
from \cref{eq:Mscrit}, where the critical temperature $T_{c0}(\mu)$ of the CFL~$\to$~NQ
transition is evaluated analytically for three massless quark flavors in the minimal
scheme. In all three cases $M_s^{\max}$ falls with $\mu$, reflecting the restoration of
chiral symmetry, while $M_{s,\text{crit}}$ rises, so that the two curves cross at a
coupling-dependent chemical potential, marked by the light-blue dashed lines in
\cref{fig:MsCrit}. To the left of the crossing the GL analysis predicts a $u$SC phase, to
the right a $d$SC phase. The comparison is meaningful, however, only where the CFL phase
melts through sequential second-order transitions, i.e., beyond the critical endpoints of
the first-order CFL phase boundary, which are indicated by the black and gray dashed
lines.

\Cref{fig:MsCrit} thus makes a GL prediction for the three couplings whether in the
RG-consistent treatment we can expect to find a melting pattern via $u$SC. 
At $G_D=G_{S}$ (right panel) the two curves cross at $\mu \approx 418\MeV$, well below the
critical points of the first-order CFL boundary at $\mu \approx 444\MeV$ and $460\MeV$. The
entire region where the GL analysis would predict a $u$SC phase is thus covered by the
first-order CFL~$\to$~2SC transition, and hence no $u$SC phase appears in the phase
diagram. 
Note, however, that in this region, at temperatures just below the first-order
phase boundary, we find the hierarchy of the gaps as $\Delta_{ud}>\Delta_{us}>\Delta_{ds}$.
If there was a second-order phase transition, we would thus expect that $\Delta_{ds}$ melts
first, leading to a $u$SC phase. It is therefore only the existence of the first-order
transition, at which $\Delta_{ds}$ and $\Delta_{us}$ jump to zero simultaneously, that
prevents this scenario from being realized.
At $G_D=\tfrac{3}{4}G_{S}$ (center panel) the crossing moves
up to $\mu \approx 452\MeV$, but the critical points, at $\mu \approx 461$ and
$470\MeV$, still lie above it, so that the potential $u$SC region remains preempted by the
first-order transition. 
However, from the trend one may expect a $u$SC phase to show up
when the diquark coupling strength is lowered further.
Indeed, at $G_D=0.6\,G_{S}$ (left panel), 
the crossing,
now at $\mu \approx 478\MeV$, 
has overtaken the critical point of the first-order CFL
boundary at $\mu \approx 452\MeV$: in the window $452\MeV \lesssim \mu \lesssim 478\MeV$
the CFL phase melts through sequential transitions \emph{and} the GL analysis predicts a
$u$SC phase. 
In this case, there must then be a doubly
critical point (DCP) \cite{Fukushima:2004zq,Abuki:2005ms,Fukushima:2005fh} in the phase diagram at which
the $u$SC phase is connected to the $d$SC phase, which exists at high densities.
As we will see in the next section, both the $u$SC phase and the DCP -- whose
location at $\mu \approx 481\MeV$ is indicated by the purple dashed line in the left panel
-- are indeed realized there.
% =====================================================================
\section{The phase diagram at \texorpdfstring{\boldmath$G_D = 0.6\,G_{S}$}{GD = 0.6 GSP}}
\label{sec:dcp}
\begin{figure}[t]
\centering
\includegraphics[width=0.95\textwidth]{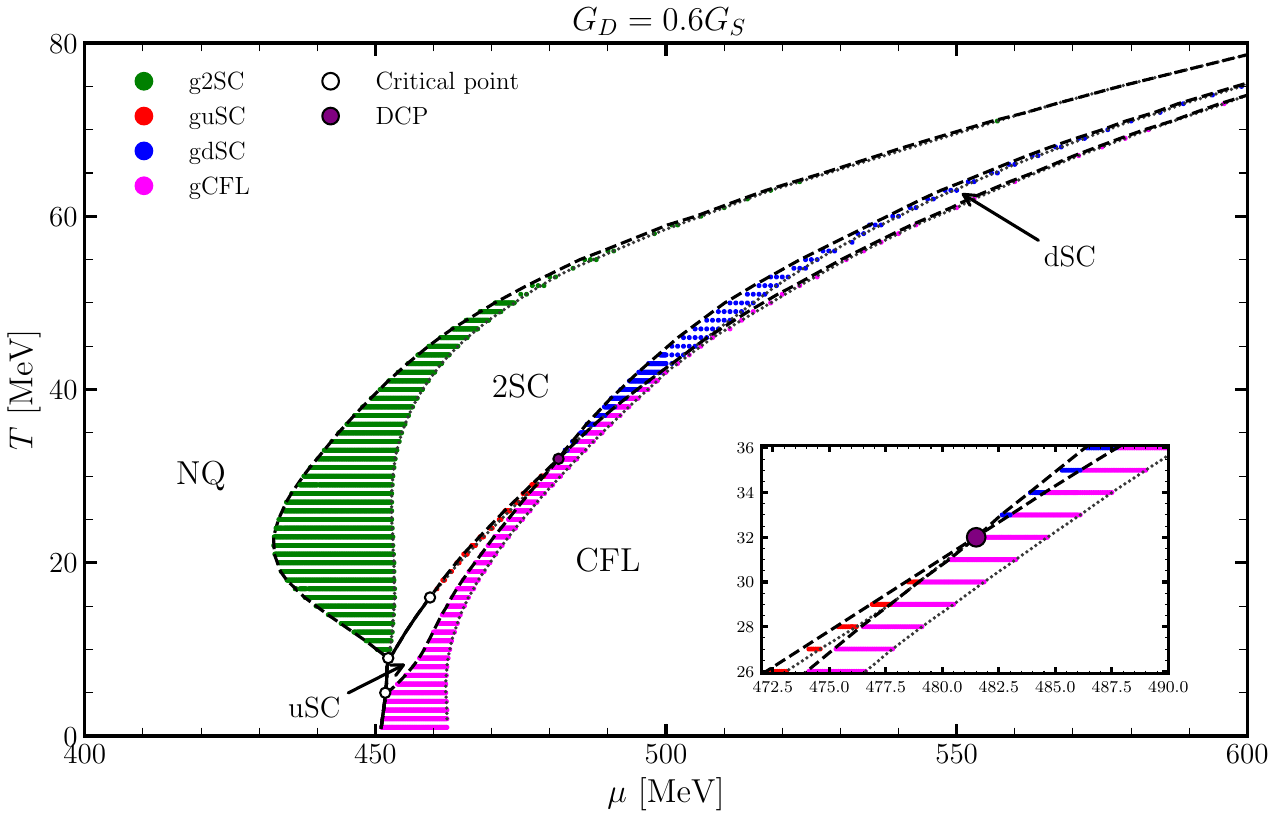}
\caption{Phase diagram of neutral quark matter in the minimal scheme at $G_D=0.6\,G_{S}$.
Gapless CSC regions are highlighted as colored hatched areas. Gray dotted lines indicate
gapless$\,\leftrightarrow\,$ungapped boundaries.
Solid (dashed) curves denote first-
(second-) order transitions. Circles mark critical points connecting first- and second- order phase boundaries. The purple dot indicates the
doubly critical point that connects four second-order lines (gCFL$\leftrightarrow$g$u$SC, g$u$SC$\leftrightarrow$ 2SC, gCFL$\leftrightarrow$g$d$SC, g$d$SC$\leftrightarrow$ 2SC).}
\label{fig:pdNJLmin06gapless}
\end{figure}

% =====================================================================
In the phase diagrams discussed so far for $G_D=G_{S}$ and $G_D=\tfrac{3}{4}G_{S}$, only a
$d$SC phase appeared. Upon lowering the coupling to $G_D=0.6\,G_{S}$, we obtain precisely the
behavior anticipated above: a DCP between $u$SC and $d$SC emerges.
\Cref{fig:pdNJLmin06gapless} displays the corresponding phase diagram, including the gapless
regions (hatched areas). 

With this small coupling the onset of color superconductivity is pushed to much larger $\mu$, and
the low-temperature part of the diagram is reorganized. At $T=0$ the system passes directly from
normal quark matter -- a gas in which the strange quark is still massive -- into CFL, more
precisely gCFL, through a first-order transition at $\mu \approx 451\MeV$, and becomes fully
gapped CFL at $\mu \approx 463\MeV$. There is no 2SC window on the $T=0$ axis at all, and in
particular no $\chi$SB$\to$2SC transition. 
This is a qualitative departure from the phase structure of
\cref{sec:gapless}, where a broad 2SC phase occupies the zero-temperature axis between the
chirally broken phase and CFL.

A related change is that the chirally broken phase no longer borders the color-superconducting
region. At $G_D=G_{S}$ and $G_D=\tfrac34 G_{S}$ there is a direct first-order transition at low temperature
from $\chi$SB into the 2SC phase or into its gapless variant, respectively. Here, at $G_D=0.6\,G_{S}$,
the pairing onset has receded to such large
$\mu$ that a wide band of normal quark matter separates the two.
The $\chi$SB--NQ
boundary is located at $\mu \approx 367\MeV$ at zero temperature and lies outside the range shown in \cref{fig:pdNJLmin06gapless}.
Chiral restoration in the
light-quark sector and the onset of color superconductivity have become two well-separated
transitions. 

Interestingly, while 2SC pairing is absent at $T=0$, for $\mu\gtrsim 430$~MeV, a g2SC phase appears when the matter is heated. For instance,  
following a vertical line at $\mu \approx 440\MeV$, the system is normal conducting at low temperature, becomes a gapless color superconductor at
$T\approx 10$~MeV and returns to normal conducting matter at $T\approx 30$~MeV. 
The mechanism, first described for neutral two-flavor quark matter in Ref.~\cite{Shovkovy:2003uu}, is the same as what we discussed before for the phase diagram at $G_D = \frac{3}{4} G_S$ with standard cutoff regularization \cite{Ruester:2005jc}: 
At $T=0$ the mismatch forbids
pairing outright; a moderate thermal smearing of the two Fermi surfaces then restores the phase
space for zero-momentum Cooper pairs without running into conflict with Pauli blocking, so that
pairing becomes possible; and only at still higher temperature do thermal fluctuations finally
destroy it.
The main difference is that in \Cref{fig:pdNJLmin06gapless} the intermediate g2SC phase does not reach all the way down to the first-order phase boundary to the $\chi$SB phase. It is possible that this happens at some slightly higher coupling $0.6 < G_D/G_S < 0.75$.

% =====================================================================
\section{Conclusion}
% =====================================================================
Enforcing RG consistency in the three-flavor NJL model removes the cutoff artifacts of the
conventional regularization and, in doing so, makes the model usable at energy scales far beyond its nominal
cutoff. A more subtle outcome is that for a sufficiently large diquark coupling, the CFL melting pattern changes from the spurious $u$SC-type pattern found in earlier self-consistent mean-field studies to the $d$SC-type pattern expected from Ginzburg--Landau analyses at
small $M_s^2/\mu^2$ \cite{Iida:2003cc,Iida:2004cj}. Investigating the phase
diagram at a smaller coupling in the minimal scheme, $G_D=0.6\,G_{S}$, we obtain the anticipated $u$SC--$d$SC doubly critical point \cite{Fukushima:2004zq,Abuki:2005ms,Fukushima:2005fh}.  The phase structure
at this coupling is markedly richer, and our RG-consistent treatment reproduces the qualitative
features reported in those earlier investigations.

It is worth emphasising why the DCP is not present the conventional treatment. The reason is
not that it lies out of reach: at $\mu \approx 480\MeV$ the DCP sits well below the nominal
cutoff $\Lambda \simeq 600\MeV$. Nor is it a matter of divergences as with a shared three-momentum cutoff, every
integral in the conventional regularization is finite. The point is what that shared cutoff removes.
The medium contribution of the paired modes falls off only like $1/p$ at large momenta, so
that it depends logarithmically on the scale up to which the modes are integrated -- the same
logarithm that turns into the divergence
$\mu^2_{\alpha a,\beta b}|\Delta_{\alpha a,\beta b}|^2\ln\Lambda$ once that scale is sent to
infinity. Cooper pairing therefore draws on modes with momenta much larger than the Fermi momentum, and truncating
all integrals at $\Lambda \simeq 600\MeV$ discards part of the pairing strength at \emph{every}
chemical potential, not only where $\mu$ approaches $\Lambda$; the loss merely worsens as the
Fermi momentum closes in on the cutoff. Two consequences remove the DCP in the conventional regularization specifically. The
first is the melting pattern: no matter the coupling, the conventional regularization yields an intermediate $u$SC phase in the temperature region above the CFL phase and never a $d$SC phase
\cite{Ruester:2005jc,Gholami:2024diy}, so there is nothing for a $u$SC region to terminate on.
The second concerns the gaps themselves:
Conventional regularization suppresses the condensates,
and the suppression gets stronger with increasing chemical potential.
On the other hand, for small diquark couplings the color-superconducting regime is shifted to higher chemical potentials.  
The conditions for which the DCP exists are thus precisely the conditions that the conventional
regularization handles worst.
In fact, if the coupling constant is too low, there are no neutral CSC phases at all with conventional regularization \cite{Duarte:2018kfd}.   

Whether such small diquark couplings are realized in nature is an open question. Astrophysical
constraints obtained within the same RG-consistent framework favor rather large couplings,
$G_D \gtrsim G_S$ \cite{Gholami:2024ety,Christian:2025dhe}, which would disfavor both the DCP scenario found here and the possibility for gapless phases at lower temperatures. On
the other hand, scalar diquarks have recently been studied with ab-initio functional QCD
methods \cite{Gholami:2026cgd}, which in principle allow to extract the model parameters,
in particular the diquark coupling, directly from QCD. Future developments along these lines
could thus determine the low-energy constants of the model and thereby help clarify whether
gapless phases or exotic structures such as the DCP can be realized in dense quark matter.

The gapless phases also point to the natural next step. Gapless color superconductors are
well known to be prone to chromomagnetic instabilities, at least at low $T$: the Meissner masses of several gluons
become imaginary in the g2SC phase \cite{Huang:2004bg,Huang:2004am}, which is usually taken as
evidence that the true ground state is inhomogeneous~\cite{Giannakis:2004pf,Kiriyama:2006xw}. 
Since
the gapless domains found here are obtained
within a framework that has removed the cutoff artifacts, it is worth asking whether they are
stable in this sense. 
In particular this concerns the region around the DCP, as three of the four phases which meet there are gapless.
Future work will address this by computing the gluon screening masses
in the paired phases directly within the RG-consistent setup. The gluon polarization tensor in
the paired medium carries the same $\mu$-dependent divergences that we identified in the
effective action, so the RG-consistent regularization procedure used here carries over to it; the
corresponding formalism is set up in Ref.~\cite{Gholami:2026qpi}. A stability analysis along
these lines, and the comparison with crystalline phases that it invites, is left for future
work.

% ---------------------------------------------------------------------
\section*{Acknowledgements}
M.B. thanks A.G.~Grunfeld for inviting him to Buenos Aires and the organizers of MAGIC 2025 for financial support.
This work was supported by the Deutsche Forschungsgemeinschaft (DFG, German Research
Foundation) -- project number 315477589 -- TRR 211. M. H. is supported by the GSI F\&E.

\section*{Data Availability}
\noindent
The numerical data presented in all figures in this work are openly
available in the ancillary files of the corresponding arXiv
submission.
The results presented in this work can be reproduced using the open-source NJL module \cite{hofmann_2026_18249033}, which was published as part of the most recent version of \textit{Calliope}, the MUSES Calculation Engine~\cite{Jahan:2026hvs}.

\bibliography{bib}

\end{document}